\documentclass[useAMS,usenatbib]{mn2e}
\usepackage{graphicx}

\title[A non-LTE abundance analysis of ROA\,5701]{A non-LTE abundance analysis of the post-AGB star ROA\,5701}

\author[H. M. A. Thompson et al.]
         {H. M. A. Thompson\thanks{email: H.Thompson@qub.ac.uk}, 
	 F. P. Keenan, P. L. Dufton, R. S. I. Ryans and J. V. Smoker
       \\
       Astrophysics Research Centre, Department of Physics \& Astronomy, The Queen's University of Belfast, Belfast BT7 1NN}

\date{Accepted 2006 February 22.  
	Received 2006 February 16;
	in original form 2005 November 16}

\pagerange{\pageref{firstpage}--\pageref{lastpage}} \pubyear{0000}
 
\def\LaTeX{L\kern-.36em\raise.3ex\hbox{a}\kern-.15em
    T\kern-.1667em\lower.7ex\hbox{E}\kern-.125emX}

\begin{document}

\label{firstpage}

\maketitle

\begin{abstract}
An analysis of high-resolution Anglo-Australian Telescope (AAT)/ University College London \'{E}chelle Spectrograph (UCL\'{E}S) optical spectra for the ultraviolet (UV)-bright star ROA\,5701 in the globular cluster $\omega$ Cen (NGC\,5139) is performed, using non-local thermodynamic equilibrium (non-LTE) model atmospheres to estimate stellar atmospheric parameters and chemical composition. Abundances are derived for C, N, O, Mg, Si and S, and compared with those found previously by Moehler et al. 
We find a general metal underabundance relative to young B-type stars, consistent with the average metallicity of the cluster.
Our results indicate that ROA\,5701 has not undergone a gas--dust separation scenario as previously suggested. However, its abundance pattern does imply that ROA\,5701 has evolved off the AGB prior to the onset of the third dredge-up. 
\end{abstract}

\begin{keywords}%
stars: abundances -- stars: AGB and post-AGB -- stars: early-type -- stars: individual: Cl$^\ast$\,NGC\,5139\,WOR\,1957 -- globular clusters: individual: NGC\,5139 
\end{keywords}

\section{Introduction}

Stars with initial masses between 0.8 and 8 M$_{\sun}$ pass through the Asymptotic Giant Branch (AGB) phase of evolution, and undergo significant mass loss \citep{ibe83}. Evolving from the tip of the AGB to the planetary nebula stage, stars move through the post-AGB phase (typically 10$^{3}$ -- 10$^{5}$ years), during which the circumstellar dust shell moves away from the star and decreases in temperature, from $\sim$ 400 to 100 K \citep{zij92}. 
The properties of post-AGB stars have been reviewed by, for example, \citet{kwo93} and \citet{van03}. This evolutionary phase is important as it provides an insight into nucleosynthesis and mixing processes during the later stages of stellar evolution. Most post-AGB stars have been identified from their infrared excess (e.g., from {\it IRAS} observations), originating from the circumstellar dust shell. This shell is more readily detected during the initial stages of post-AGB evolution, when it is relatively hot and close to the star, and hence the majority of stars identified as being at a post-AGB evolutionary stage have just left the AGB and are of spectral type A, F or G \citep{oud96}. However, there are many features of post-AGB evolution which are not yet fully understood. For example, several stars exhibit abundance patterns similar to interstellar gas, with elements such as Mg, Si, Al, Ca, Ti etc following the very low Fe abundance, while C, N, O and S are near solar (\citealt*{van95}; see, for example, \citealt*{{luc84},{par92},{nap94}}). 
\citet{mat92} discussed the conditions necessary for such a `cleaned-up' photosphere in post-AGB stars, suggesting that the process is likely to occur during mass transfer between members of a binary system, or accretion triggered by a binary companion. \citet*{wat92} added that accretion of material from a circumstellar disc around a binary system was a favorable scenario. In this, grains form in the circumstellar shell produced during the AGB phase, and the subsequent slow accretion of gas from the material results in the observed abundance patterns. \citet{van95} found that all the extremely iron deficient post-AGB stars, known at that time, were binaries, supporting this hypothesis.

Although the majority of post-AGB candidates are cool objects, several B-type stars originally classified as Population {\sc i} stars are now believed to be at this evolutionary stage, on the basis of their derived atmospheric parameters and chemical compositions (see, for example, \citealt{rya03}). These B-type stars should be evolving from the cooler post-AGB stars, but significant differences are displayed. The most obvious is the large carbon underabundance found for several hot objects, which may imply that they have left the AGB prior to the onset of the third dredge-up (for example, \citealt*{{con91},{mcc92},{rya03},{jas04}}). 

The abundance differences between hot and cool post-AGB candidates indicate that further study is required to fully understand this evolutionary stage. It is important to begin by studying hot post-AGB objects of known initial metallicity, rather than field stars of unknown origin. Such a sample are the ultraviolet (UV)-bright stars found in several globular clusters \citep*{{con94},{dix04},{moo04}}, whose post-AGB nature has been established from their position in the cluster colour--magnitude diagram \citep{lan00}. The term `UV-bright' was first suggested by \citet*{zin72} and denotes stars which have ultraviolet fluxes larger than those in either the red giants or the horizontal branch stars \citep{nor74}. For an excellent review of hot stars in globular clusters see \citet{moe01}. 

ROA\,5701, in the Galactic globular cluster $\omega$ Centauri, is a well studied UV-bright star due to its relative brightness (\citealt*{{nor74},{cac84},{moe98}}). The metal-poor $\omega$ Cen is known to have (at least) three stellar populations, with metallicities: [Fe/H] $\sim$ --1.6, corresponding to the metal poor population containing the bulk of the stars, [Fe/H] $\sim$ --1.2 for an intermediate metallicity population and [Fe/H] $\sim$ --0.5 for a metal-rich population comprising 5 per cent of the stars (\citealt*{{bed04},{hil04},{nor04},{pio05}}). 
In this paper, we analyse high-resolution optical spectra of ROA\,5701 using non-local thermodynamic equilibrium (non-LTE) model atmosphere techniques to estimate its atmospheric parameters and chemical composition, leading to a reconsideration of its evolutionary status.

\section{Observations and data reduction}

The spectroscopic data presented here were obtained with the 3.9-m Anglo--Australian Telescope (AAT) on the nights of 2000 July 15 and 16. The University College London \'{E}chelle Spectrograph (UCL\'{E}S) was used with the MITLL 2A CCD (2048 $\times$ 4096) detector, the 31 groove ${\rm mm^{-1}}$ grating and the 70 cm camera, giving a linear dispersion of 2 \AA\,${\rm mm^{-1}}$ and a spectral resolution [full width at half maximum (FWHM)] of {\it R} $\sim$ 40000. The read-out noise and gain were 1.8 ${\rm e^-}$ and 0.37 ${\rm e^-\,ADU^{-1}}$ respectively. Two overlapping wavelength regions were observed, resulting in a spectral coverage of 3760--9400 \AA{}. Flat-field and bias frames were exposed at the beginning and end of the nights, while thorium-argon wavelength calibration spectra were interweaved with the stellar spectra. 
Reduction of the raw two dimensional CCD images were performed using standard procedures within the Image Reduction and Analysis Facility ({\sc iraf}; \citealt{bar93}). Procedures such as bias subtraction and flat fielding were achieved using the {\sc ccdred} package \citep{mas97}. Cosmic-ray removal, sky subtraction, stellar extraction and wavelength calibration were performed using {\sc specred} \citep*{mas92} and {\sc doecslit} \citep{val93}.

\begin{figure}
\includegraphics[angle=270,width=0.5\textwidth]{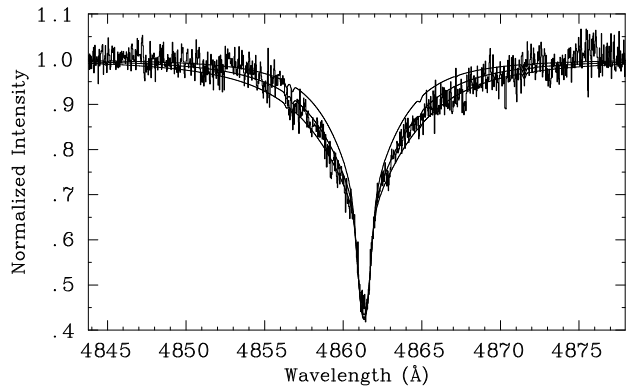}
\caption{Observed and theoretical line profile for H$\beta$ at 4861 \AA{} in the spectrum of ROA\,5701. The theoretical profiles have been generated for surface gravities of 3.0, 3.25 and 3.5 dex (lower, middle and upper smooth curves, respectively), and imply a surface gravity of log~{\em g} = 3.2 dex.}
\label{fig_H}
\end{figure}

After extraction, the spectra were co-added (resulting in a signal-to-noise ratio of $\sim$ 30) and inserted into the Starlink package {\sc dipso} \citep{how04} for further analysis. The spectral continuum was normalised, with low-order ($<$ 5) polynomials in regions of spectra free from absorption lines. Following this, they were radial velocity shifted to the restframe, using unblended metal and hydrogen lines. This resulted in the kinematical Local Standard of Rest radial velocity ${\rm v_{lsr}}$ = +246 $\pm$ 3 ${\rm km s^{-1}}$, using corrections generated by the Starlink progam {\sc rv} \citep{wal96}. Equivalent widths were measured for metal absorption lines by utilising the emission line fitting ({\sc ELF}) routines within {\sc dipso}. {\sc ELF} uses non-linear least squares Gaussian fitting, with line--centre, linewidth and line--strength as variable parameters. The typical FWHM for the metal absorption line spectrum was $\sim$ 11 ${\rm km s^{-1}}$, indicating that a small amount of broadening is evident in the lines. 
The error associated with the measurement of equivalent widths has been previously discussed \citep{duf90}, with a well-observed unblended feature having an accuracy of typically 10 per cent (designated as quality a in Table \ref{tab_all}), a weak or blended feature accurate to 20 per cent (designated b), and weaker features with accuracy less than 20 per cent (designated c). For the Balmer hydrogen lines (i.e., H$\alpha$, H$\beta$, H$\delta$, H$\gamma$) and diffuse He\,{\sc i} lines, equivalent widths were not measured, but theoretical profiles were fitted to the normalized observational data (see, for example, Fig. \ref{fig_H}).

\begin{table}
\begin{center}
\caption{Equivalent width measurements and corresponding abundances for the observed metal lines.}
\label{tab_all}
\begin{tabular}{@{}lcccccccc}
Wavelength (\AA) &Species 	& W (m\AA)	& Quality$^\dagger$ & Abundance$^\star$ \\
\hline 
3911.96$^+$ 	& O\,{\sc ii}	&	23 	     & a     & 7.52  \\
3919.28 	& O\,{\sc ii} 	&	30 	     & b     & 7.98  \\
3945.04 	& O\,{\sc ii} 	& 	17 	     & b     & 7.65  \\
3954.37 	& O\,{\sc ii} 	& 	41 	     & a     & 7.81  \\
3982.71 	& O\,{\sc ii}	& 	19 	     & c     & 7.70  \\
4069.62$^{+\ast}$	& O\,{\sc ii} 	& 	42 	     & a     & 7.72  \\
4069.89$^{+\ast}$ 	& O\,{\sc ii} 	& 	47 	     & a     & $\cdots$  \\
4072.16 	& O\,{\sc ii} 	& 	59	     & a     & 7.60  \\
4075.86 	& O\,{\sc ii} 	& 	66	     & a     & 7.52  \\
4078.84 	& O\,{\sc ii} 	& 	18 	     & a     & 7.70  \\
4132.80 	& O\,{\sc ii} 	& 	27 	     & a     & 7.79  \\
4185.46 	& O\,{\sc ii} 	& 	31 	     & a     & 7.77  \\
4317.14$^+$ 	& O\,{\sc ii} 	& 	38	     & a     & 7.75  \\
4319.63 	& O\,{\sc ii} 	& 	36 	     & a     & 7.71  \\
4349.43 	& O\,{\sc ii} 	& 	64 	     & a     & 7.73  \\
4351.27$^+$ 	& O\,{\sc ii} 	& 	38 	     & a     & 7.46  \\
4366.90$^+$ 	& O\,{\sc ii} 	& 	40 	     & a     & 7.71  \\
4369.27 	& O\,{\sc ii} 	& 	11 	     & c     & 7.83  \\
4395.93 	& O\,{\sc ii} 	& 	16 	     & c     & 7.84  \\
4414.91 	& O\,{\sc ii} 	& 	64 	     & a     & 7.67  \\
4416.98 	& O\,{\sc ii} 	& 	50	     & a     & 7.77  \\
4590.97 	& O\,{\sc ii} 	& 	45 	     & a     & 7.61  \\
4596.17 	& O\,{\sc ii} 	& 	40 	     & a     & 7.65  \\
4638.86 	& O\,{\sc ii} 	& 	52 	     & a     & 7.82  \\
4641.81 	& O\,{\sc ii} 	& 	87 	     & a     & 7.79  \\
4649.14$^{+\ast}$	& O\,{\sc ii} 	& 	94 	     & a     & 7.67  \\
4650.84$^{+\ast}$	& O\,{\sc ii}	& 	47 	     & a     & $\cdots$  \\
4661.64 	& O\,{\sc ii} 	& 	50 	     & a     & 7.72  \\
4676.24 	& O\,{\sc ii} 	& 	39	     & b     & 7.69  \\
4699.22$^+$ 	& O\,{\sc ii} 	& 	28 	     & b     & 7.55  \\
4705.35 	& O\,{\sc ii} 	& 	32	     & b     & 7.59  \\
4710.01$^+$ 	& O\,{\sc ii} 	& 	20 	     & b     & 8.17  \\
4906.83 	& O\,{\sc ii} 	& 	18 	     & c     & 7.62  \\
4924.53 	& O\,{\sc ii} 	& 	38 	     & a     & 8.00  \\
4941.07 	& O\,{\sc ii} 	& 	16 	     & b     & 8.01  \\
4943.00 	& O\,{\sc ii} 	& 	20 	     & b     & 7.89  \\
3995.00$^+$ 	& N\,{\sc ii} 	& 	44	     & b     & 6.97  \\
4241.78$^+$ 	& N\,{\sc ii} 	& 	14	     & c     & 7.57  \\
4447.03 	& N\,{\sc ii} 	& 	12 	     & c     & 6.81  \\
4630.54 	& N\,{\sc ii} 	& 	30 	     & b     & 7.00  \\
4643.09 	& N\,{\sc ii} 	& 	17	     & b     & 7.17  \\
5001.00$^+$	& N\,{\sc ii} 	&	46	     & b     & 6.92  \\
5005.15		& N\,{\sc ii} 	&	31	     & b     & 6.88  \\
5007.33 	& N\,{\sc ii} 	&	14	     & c     & 7.06  \\
4267.00$^+$ 	& C\,{\sc ii} 	& 	$<$ 10 	     & $\cdots$  & $<$ 6.28  \\
6578.05 	& C\,{\sc ii} 	&	$<$ 10	     & $\cdots$  & $<$ 6.51  \\
4088.85 	& Si\,{\sc iv} 	& 	38	     & a     & 5.97  \\
4116.10 	& Si\,{\sc iv} 	& 	22	     & a     & 5.91  \\
4552.62 	& Si\,{\sc iii} & 	73 	     & a     & 5.94  \\
4567.82 	& Si\,{\sc iii} & 	50 	     & a     & 5.93  \\
4574.76 	& Si\,{\sc iii} & 	22	     & a     & 5.95  \\
4481.00$^+$ 	& Mg\,{\sc ii} 	& 	21	     & c     & 6.27  \\
4253.49 	& S\,{\sc iii} 	& 	17 	     & c     & 5.71  \\
4529.19 	& Al\,{\sc iii} & 	$<$ 10	     & $\cdots$  & $<$ 5.21  \\
\hline
\end{tabular}
\end{center}
\medskip
$^\dagger$a = a typical accuracy of 10 per cent, b = a typical accuracy of 20 per cent, c = an accuracy less than 20 per cent.\\
$^\star$Logarithmic abundance [M/H] on the scale log[H] = 12.00.\\
$^+$Treated as blends in {\sc tlusty}.\\
$^\ast$The equivalent widths of the adjacent lines were combined to produce the corresponding abundances shown.
\end{table}

\section{Analysis}

\subsection{Non-LTE atmosphere calculations}

Non-Local Thermodynamic Equilibrium (non-LTE) model atmosphere grids have been calculated using the codes {\sc tlusty} and {\sc synspec} \citep*{{hub88},{hub95},{hub98}} and employed to derive chemical abundances and atmospheric parameters. Detailed discussions of the grids and methods can be found in \citet{rya03} and \citet{duf05}, and also at http://star.pst.qub.ac.uk . 

Essentially, four grids have been developed with metallicities based on our Galaxy (i.e., Fe abundance of [Fe/H] = 7.5 dex, where [H] = 12.00), and metallicites reduced by 0.3 dex [Large Magellanic Cloud (LMC)], 0.6 dex [Small Magellanic Cloud (SMC)] and 1.1 dex (for low metallicity regimes).
Approximately 3000 models were calculated per metallicity (i.e., 12000 in total) for a range of effective temperatures ($T_{\rm eff}$ = 12000--35000 K), logarithmic gravities (log~{\em g} = 4.5 dex down to close to the Eddington limit depending on the effective temperature) and microturbulences ($\xi$ = 0--30 ${\rm km\, s^{-1}}$). For each grid the iron abundance was fixed and the other abundances (C, N, O, Mg, Si, S) were varied around their base rate from --0.8 to +0.8 dex. 

A crucial assumption with this method is that the metal line blanketing is dominated by iron. The atmospheric structure is then defined by the iron abundance (metallicity), the effective temperature, gravity and microturbulance (which affects the amount of blanketing). 
\citet{hub98} have discussed the effects of including different amounts of line blanketing in their {\sc tlusty} model atmospheres. They find that while the inclusion of light elements is important, the iron group elements dominate the line blanketing. Additionally a comparison for models including only iron, and iron and nickel, line blanketing indicated that the inclusion of nickel leads to only modest changes with, for example, the UV continuum changing by about 5 per cent. Hence although the exclusion of this source of blanketing may lead to additional uncertainty in the atmospheric parameters and chemical compositions, it is unlikely to invalidate the results found here. 
It is also assumed that the adopted abundances of the light elements can vary without significantly affecting the atmospheric structure. \citet{duf05} tested this assumption by considering a variety of sets of atmospheric parameters. For each set, the abundance of one of the light elements was fixed, while those of the other light elements were allowed to vary. They found that changes in equivalent widths were always less that 5 per cent, thus implying that the assumption is reasonable. 

The models were used to calculate spectra, which provide theoretical profiles of H\,{\sc i} and He\.{\sc i} lines, and profiles and equivalent widths of transitions in light metals for a range of abundances. The theoretical metal line equivalent widths were accessed via a {\sc GUI} interface written in Interactive Data Language ({\sc IDL}), which allows the user to interpolate in order to calculate equivalent widths and abundance estimates for approximately 200 metal lines for any given set of atmospheric parameters. \citet{rya03} reported that the abundance increment of 0.4 dex used in the grids was sufficiently fine to ensure that no significant errors were introduced by the interpolation procedures. Full theoretical spectra are also available for any given model.

\subsection{Atmospheric parameters}

The atmospheric parameters were determined using standard procedures discussed in, for example, \citet*{{kil92},{tru04},{hun05}}, and will only be covered here briefly. The parameters of effective temperature, gravity, microturbulence and metallicity (iron content) of the star are interrelated so an iterative process is required to determine each parameter. The grid with an iron abundance of 1.1 dex below solar was selected, as this was most appropriate to the metallicity of the $\omega$ Cen globular cluster, namely [Fe/H] = --1.2 dex \citep{pio05}. The validity of this grid selection will be discussed below. The adopted atmospheric parameters and associated errors are listed in Table \ref{tab_ap}, together with those derived previously for ROA\,5701 \citep{{nor74},{cac84},{moe98}}.

\begin{table}
\begin{center}
\caption{Adopted atmospheric parameters for ROA\,5701 as derived here and previously by \citet{nor74}, \citet{cac84} and \citet{moe98}.}
\label{tab_ap}
\begin{tabular}{@{}lccccccc}
\hline
Star		& $T_{\rm eff}$		& log~{\em g}		& $\xi$	\\
		& ${\rm 10^{3}\,K}$	& dex			& ${\rm km\, s^{-1}}$ \\
\hline
This paper	& 25 $\pm$ 1		& 3.2 $\pm$ 0.2		& 5 $\pm$ 5	\\
\citet{moe98}	& 22--24.5		& 3.2--3.4		& 2--3 or 20	\\
\citet{cac84}	& 24			& $\cdots$		& $\cdots$	\\
\citet{nor74}	& 26.6			& 3.5			& $\cdots$	\\
\hline
\end{tabular}
\end{center}
\end{table}

\subsubsection{Effective temperature, $T_{\rm eff}$}

An estimate for the effective temperature, $T_{\rm eff}$, was determined by balancing the abundance estimated for silicon from different ionization stages (Si\,{\sc iii} and Si\,{\sc iv}), initially assuming a surface gravity of 3.2 dex and a microturbulence of 5 ${\rm km\, s^{-1}}$, leading to a final effective temperature estimate, $T_{\rm eff}$ = 25000 K. The He\,{\sc ii} line at 4541.59 \AA {} was not observed, and using theoretical profiles it was possible to put an upper limit of $T_{\rm eff}$ $<$ 26000 K, assuming a normal helium abundance. Therefore, the error adopted for the effective temperature is $\pm$ 1000 K, which would produce a change in the silicon abundance estimates of $\pm$ 0.1 dex for Si\,{\sc iii} and $\pm$ 0.3 dex for Si\,{\sc iv}. 
\citet{lee05} found that using grids of different iron abundances changed the estimate by typically $\leq$ 500 K, so this should not be a major source of error.

\begin{figure}
\includegraphics[angle=270,width=0.5\textwidth]{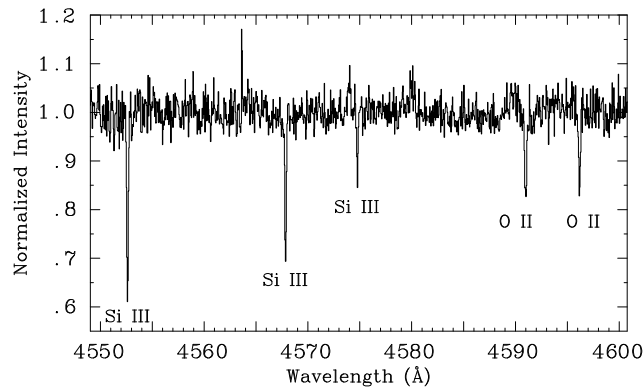}
\caption{Portion of the observed spectrum for ROA\,5701 in the restframe, covering the wavelength region 4550--4600 \AA, showing several O\,{\sc ii} and Si\,{\sc iii} transitions.}
\label{fig_si}
\end{figure}

\subsubsection{Logarithmic gravity, log~{\em g}}

The surface gravity was estimated by comparing the observed spectrum in the region of the Balmer lines with theoretical profiles, and led to an estimate of log~{\em g} = 3.2 dex (Fig. \ref{fig_H}). Using different iron abundance grids did not affect the gravity estimate, while the quality of the observed spectrum implies an uncertainty associated with the fitting of approximately $\pm$ 0.2 dex.

\subsubsection{Microturbulence, $\xi$}

The microturbulence was initially assumed to be 5 ${\rm km\, s^{-1}}$ for the purposes of estimating the effective temperature and surface gravity. In order to constrain this value, the abundance estimates from 34 observed O\,{\sc ii} lines were initially assumed to be independent of the line strength (i.e., a plot of abundance against line strength has a zero gradient). However, \citet{hun05} state that using O\,{\sc ii} lines from different multiplets may compromise the validity of this method. Therefore, the Si\,{\sc iii} multiplet (Fig. \ref{fig_si}) was also used to remove this compromise. 
The O\,{\sc ii} and Si\,{\sc iii} multiplets were in agreement, giving an estimate of 5 ${\rm km\, s^{-1}}$, with an adopted error of $\pm$ 5 ${\rm km\, s^{-1}}$ to allow for the uncertainty in the analysis. Using different metallicity grids again did not alter these estimates.

\subsection{Abundances}

The adopted atmospheric parameters (as shown in Table \ref{tab_ap}) were used to determine non-LTE abundances for ROA\,5701. The atomic data were mainly taken from http://star.pst.qub.ac.uk , with additional sources being \citet{{ham96a},{ham96b}} and \citet*{sar05}. Table \ref{tab_all} contains the observed wavelengths at the restframe, equivalent widths and abundance estimates for the metal lines observed. For the C\,{\sc ii} and Al\,{\sc iii} lines, it was only possible to derive upper limits for the equivalent widths and corresponding element abundances. These upper limits were estimated by fitting theoretical profiles to the appropriate spectral region, allowing for the quality of the spectra and surrounding features. In some cases more than one line contributes to the measured equivalent width, and these lines are treated as blends in {\sc tlusty}.

Table \ref{tab_ab} contains the abundance estimates for each species along with their error estimates. 
Also included are the abundance values found previously by \citet{moe98} for ROA\,5701. Non-LTE abundance estimates for samples of young B-type stars, both in the local field \citep{{kil92},{kil94}} and in the young Galactic cluster NGC\,6611 (Hunter; in preparation), are also presented. The latter have been determined using the {\sc tlusty} code, as in the present paper. Typical interstellar medium (ISM) abundances for the elements are also tabulated \citep*{wil00}.   

The errors associated with the derived abundances arise from random errors connected with analysing the data, such as observational uncertainties, errors in oscillator strengths and line fitting, and from systematic errors arising from the stellar atmospheric parameter estimates. The random uncertainty is taken as the standard deviation of the abundances for any given species divided by the square root of the number of lines observed of that species. If only one line of a species was observed, e.g., Mg\,{\sc ii}, the random uncertainty was taken to be the standard deviation of the best-observed species, O\,{\sc ii}. The systematic uncertainties were estimated by changing each parameter in turn by its associated error, and the resulting abundance was compared to the abundance values shown in Table \ref{tab_ab}. The random and systematic errors were then squared, summed and square rooted to give the total uncertainties as shown in Table \ref{tab_ab}. \citet{hun05} gives further details of this methodology.

There are currently no non-LTE grids available for S\,{\sc iii} and Al\,{\sc iii}, and for these elements a non-LTE model was calculated (using {\sc tlusty}) for the adopted atmospheric parameters, which was then used to generate theoretical LTE equivalent widths. These were used to derive abundance estimates, which are listed in Table \ref{tab_ab} along with the uncertainties utilizing the same method as discussed above.

\section{Results and discussion}

\begin{table*}
\begin{minipage}{\textwidth}
\begin{center}
\caption{Absolute abundances for ROA\,5701 along with the estimated uncertainties. The bracketed values (column two) represent the number of lines observed for each species. Also listed are the results derived for ROA\,5701 by \citet{moe98}, the values for young B-type stars in the local field \citep{{kil92},{kil94}} and in the Galactic cluster NGC\,6611 (Hunter; in preparation), and abundances in the interstellar medium \citep{wil00}.}
\label{tab_ab}
\begin{tabular}{@{}l ccccccccc}
\hline\\
		&\multicolumn{3}{c}{ROA\,5701}					&\multicolumn{2}{c}{Young B-type Star}	& 	\\
Species 	&\multicolumn{2}{c}{This paper}			&\citet{moe98} 	& Local Field	&NGC\,6611 	& ISM	\\
		&\multicolumn{3}{c}{}						& 		&		& 	\\
\hline \\
C\,{\sc ii} 	& $<$6.51 $\pm$ 0.18		&(1)	 	&$<$5.85	& 8.20 	& 7.95  & 8.38  \\
N\,{\sc ii} 	& 7.05 $\pm$ 0.15		&(8)	 	& 6.86  	& 7.69  & 7.59  & 7.88  \\
O\,{\sc ii} 	& 7.75 $\pm$ 0.13		&(34)	 	& 7.97 		& 8.55  & 8.55  & 8.69  \\
Mg\,{\sc ii}	& 6.27 $\pm$ 0.19		&(1)	 	& $\cdots$ 	& 7.38  & 7.32  & 7.40  \\
Si\,{\sc iii} 	& 5.94 $\pm$ 0.27		&(3)	 	& 6.14  	& 7.28  & 7.41  & 7.27  \\
Si\,{\sc iv} 	& 5.95 $\pm$ 0.47		&(2)	 	& $\cdots$	& $\cdots$&$\cdots$   & $\cdots$   \\
Al\,{\sc iii} 	& $<$5.21 $\pm$ 0.23		&(1)	 	& $\cdots$	& 6.22  & $\cdots$   & 6.33   \\
S\,{\sc iii} 	& 5.71 $\pm$ 0.21		&(1)	 	& $\cdots$ 	& 7.25  & $\cdots$   & 7.09  \\
Fe\,{\sc iii} 	&       $\cdots$		&	 	& 4.79  	& $\cdots$& 7.50  & 7.43  \\
\hline
\end{tabular}
\end{center}
\end{minipage}
\end{table*}

The adopted atmospheric parameters are shown in Table \ref{tab_ap}. The absolute abundance estimates are summarized in Table \ref{tab_ab}, with differential abundances in Table \ref{tab_ay}. 
In the following subsections these results are examined, and considered with previous work, leading to a discussion of the evolutionary status of ROA\,5701.

\subsection{Chemical composition of ROA\,5701}

Helium abundances have not been derived explicitly in our work; instead a normal helium abundance has been assumed throughout the analysis. To test this assumption, theoretical model profiles were compared with the observed spectra at 4026, 4471 and 4921 \AA. Within the uncertainties of the atmospheric parameters, there is good agreement between theory and observation, and therefore this assumption appears valid.

There are a number of absorption lines in the ROA\,5701 spectrum due to O\,{\sc ii} and N\,{\sc ii}, for the majority of which the accuracy of their equivalent widths was better than 20 per cent. For oxygen there are four lines, and nitrogen three lines, with an accuracy worse than 20 per cent. If these lines are not included in the absolute abundance estimates, the resulting values would still be within the uncertainty of the abundances listed in Table \ref{tab_ab}. Therefore, including lines of poorer observational quality does not seem to adversely affect the abundance estimates. Although other lines attributed to O\,{\sc ii} were detected, they are blended with hydrogen and helium lines and have not been included in the non-LTE calculations. 

The carbon abundance is given as an upper limit, as it was only possible to estimate an upper limit for the C\,{\sc ii} line equivalent widths (at 4267.00 and 6578.05 \AA), due to the quality of the spectra. \citet{ebe88} found that a carbon abundance from the 4267-\AA{} line can be reduced by up to 0.5 dex due to non-LTE effects. \citet{sig96} demonstrated, through detailed non-LTE calculations, that the 6578-\AA{} line is more reliable than the 4267-\AA{} line for $T_{\rm eff}$ $<$ 25000 K. Therefore, the abundance shown in Table \ref{tab_ab} is from the 6578-\AA{} line only. 

For magnesium, it was possible to observe the doublet of Mg\,{\sc ii} at 4481 \AA, but the abundance appears reliable. The abundance estimates from the Si\,{\sc iii} spectrum are less sensitive to changes in atmospheric parameters compared with those from Si\,{\sc iv}. As previously discussed, a change of effective temperature ($\pm$ 1000 K) alters the Si\,{\sc iii} abundance by $\pm$ 0.1 dex and the Si\,{\sc iv} abundance by $\pm$ 0.3 dex, which is reflected in the error estimates in Table \ref{tab_ab}. 

A sulphur abundance was derived from a single line at 4253.49 \AA, which is not normally used, due to the difficulty in resolving it from O\,{\sc ii} lines at 4253.89 and 4253.91 \AA, for stars with significant rotational velocities. To investigate that this is indeed real and not a noise feature, the spectrum of $\gamma$ Peg, a main--sequence B-type star of $T_{\rm eff}$ = 24000 K and log~{\em g} = 4.0 dex \citep{rya96}, was examined. This confirmed that the 4253.49-\AA{} line is stronger than other observed sulphur features (i.e., those at 4284.90, 4361.47 and 4364.66 \AA). The equivalent width (and hence abundance) derived does not appear to be affected by any contributions from the O\,{\sc ii} lines, although the abundance obtained is based on LTE calculations and may not be as reliable as those from a non-LTE analysis. However, \citet{hun05} found that for sulphur, the mean LTE abundance was lower than the non-LTE abundance by only 0.01 dex, and so concluded that non-LTE effects appear negligible for this species when using the appropriate atmospheric parameters.

Aluminium was not observed so an upper limit was set for the equivalent width and abundance estimate using the Al\,{\sc iii} line at 4529.19 \AA. This abundance in ROA\,5701 is also based on LTE calculations and should be treated with caution. However, \citet{hun05} state that if the dominant ion stage is used to calculate the LTE abundance then this value will generally be in good agreement with non-LTE abundances.  
A feature was observed at 4512.41 \AA{} and had been originally classified as Al\,{\sc iii} (at 4512.56 \AA). However, using the $\gamma$ Peg spectrum, the relative line strengths were examined, showing that the Al\,{\sc iii} line at 4529.19 \AA{} should be a stronger feature. 
There was no absorption visible at 4529.19 \AA{} indicating that the line at 4512.41 \AA{}, if real, is not due to Al\,{\sc iii}. 

No absorption due to iron was detected in the spectrum of ROA\,5701. Adopting the average metallicity of the cluster (i.e., [Fe/H] = --1.2 dex; \citealt{pio05}), synthetic spectra were derived to investigate if strong Fe\,{\sc iii} features, e.g., at 4419.60 \AA{}, would be observable. It was noted that, with spectra of a higher signal-to-noise, it is possible that these features would be observable, with predicted equivalent widths of $\ga$ 10 m\AA{}. There was another unidentified feature, at 4352.55 \AA, which was originally thought to be a Fe\,{\sc iii} line at 4352.58 \AA. However examining the spectrum of $\gamma$ Peg, the line strengths for other known Fe\,{\sc iii} lines (i.e., at 4164.73, 4419.60 and 4431.02 \AA) were found to all be stronger than the line at 4352.58 \AA, also seen in $\gamma$ Peg. It is unlikely that the observed feature is due to iron as there are no other absorption lines of Fe\,{\sc iii} detected. Another possibility is that is is due to nitrogen, as there is a N{\sc ii} line expected at 4352.22 \AA. However, model atmospheric calculations indicate that this N\,{\sc ii} feature should be very weak and so no identification for this line is currently available.

\subsection{Comparisons with previous work}

\subsubsection{Atmospheric parameters}

In Table \ref{tab_ap} the atmospheric parameters for ROA\,5701 are summarized as derived in this paper and from previous studies \citep{{nor74},{cac84},{moe98}}. From an analysis of narrow--band and {\it UBV} observations, \citet{nor74} found $T_{\rm eff}$ and log~{\em g} estimates. \citet{cac84} obtained an estimate for the effective temperature based on UV spectra from the {\it International Ultraviolet Explorer}({\it IUE}) satellite. \citet{moe98} used high resolution ESO Cassegrain Echelle Spectrograph (CASPEC) optical spectra and UV data obtained with the Goddard High Resolution Spectrograph (GHRS) to obtain their results. 

Comparing the atmospheric parameters obtained in this paper, indicates that there is good agreement with those previously found. 
The effective temperature and surface gravity estimates are within the ranges of the previous values, with the exception of \citet{nor74}. This is most likely due to different values of ({\it B--V}) used in their calculations, as cited in \citet{cac84}. \citet{kil92} states that a spectroscopic analysis produces more reliable atmospheric parameters than a photometric analysis, so the Balmer line profiles and the silicon ionization equilibrium should be reliable methods for $T_{\rm eff}$ and log~{\em g}, indicating that the values deduced here are credible. 

The microturbulence obtained by \citet{moe98} differs depending on the method adopted. Two values are presented, the larger (20 ${\rm km\, s^{-1}}$) derived from the O\,{\sc ii} lines, but if Fe\,{\sc iii} transitions are employed the value decreased to 2--3 ${\rm km\, s^{-1}}$. 
\citet{gie92} note that including non-LTE effects reduces high microturbulence velocities compared to using LTE effects, from 20 ${\rm km\, s^{-1}}$ to approximately 10 ${\rm km\, s^{-1}}$ for the O\,{\sc ii} spectrum. 
We test this for our data using the {\sc tlusty} grid to obtain a LTE microturbulance from our observed lines. An estimate of 10 ${\rm km\, s^{-1}}$ was calculated, which is higher than our non-LTE value, thus in agreement with \citet{gie92}. Allowing for non-LTE effects gives agreement between our values and that of \citet{moe98}, within the estimated error. In general, we believe that the values presented in this paper are more reliable than the previous results discussed, due to improvements in the quality of the observational data and the theoretical models adopted.

\subsubsection{Chemical composition}

\citet{moe98} have previously analysed ROA\,5701, and their absolute abundance estimates are shown in Table \ref{tab_ab}, along with our abundance estimates. As stated above, they used CASPEC and GHRS spectra, using both a curve-of-growth method and spectrum synthesis with an LTE grid. From the optical spectrum, abundances for C, N, O and Si were obtained, and an iron abundance from the UV spectrum; Mg, Al and S were not observed. Error estimates for the measured equivalent widths were not included in their tables 2 and 3, but for the values given there is general agreement with those found here (to within approximately $\pm$ 10 m\AA). We note that the \citet{moe98} abundance results in Table \ref{tab_ab} are the average of their curve-of-growth and spectrum sythesis results. 

Absolute abundances may contain systematic errors due to uncertainties in the atmospheric parameters and atomic data. In order to minimize such errors, the ROA\,5701 abundances were also compared with young B-type star values, of known chemical composition (Population {\sc i}), found using similar non-LTE models (Table \ref{tab_ay}, which also contains differential abundances relative to the ISM).Where possible, the results from the cluster NGC\,6611 (Hunter; in preparation) were used, as these were derived using the same model atmosphere non-LTE calculations.

Comparing our abundances with those found by \citet{moe98}, we see similar trends of underabundance. Their carbon value is much lower than ours despite having used the same upper limit value for the equivalent width. This lower abundance could be due to differences in their model atmosphere calculations, especially given the large non-LTE effects predicted for the C\,{\sc ii} ion \citep{sig96}, but the result found here is in better agreement with the C abundance of other stars found in $\omega$ Cen. 

Our O, N and Si abundance values differ from \citet{moe98} by typically 0.2 dex, this could be due to non-LTE effects. For example, \citet{gie92} and \citet{kil94} found that, for B-type stars, LTE abundances agree to 0.2 dex with corresponding non-LTE values. 
To test if non-LTE effects are the cause of abundance differences, we have derived, using the {\sc tlusty} grid, LTE abundances for carbon of $<$6.25, nitrogen of 7.12, oxygen of 7.96 and silicon of 6.32 dex. Our oxygen abundance is in very good agreement with that of \citet{moe98}. The nitrogen and silicon abundances are 0.26 and 0.18 dex, respectively, larger, while the carbon abundance is in better agreement with that of \citet{moe98}, compared to our non-LTE value, although it is still an upper limit. The increase in LTE abundance estimates for C, N and Si, compared to those of \citet{moe98}, could be due to using different atmospheric parameters and equivalent width estimates. The non-LTE results should be more reliable, as non-LTE models and more lines have been used to calculate the abundances, thereby reducing associated errors. 

No abundance estimates were derived by \citet{moe98} for Mg or S, as no lines from these elements were observed. Therefore, our results yield new abundance measurements for these species in ROA\,5701. Compared with the other derived abundances, they show similar abundance patterns and this is discussed further in the following section. 

\citet{moe98} were able to use their UV spectrum to obtain a differential iron abundance of [Fe/H] = --2.7 dex, which is lower than that found for any other star in $\omega$ Cen to date. However, even in metal-poor stars, the UV spectral region is crowded so continuum placement is difficult. \citet{moe98} indicate that an uncertainty of 5 per cent in the continuum definition of their spectrum would give an error of 0.2 dex in their iron abundance. There may also be systematic errors which could be reduced if abundances were compared with young B-type stars. This is unfortunately not possible for the UV Fe\,{\sc iii} lines, as normal B-type stellar spectra are affected by line crowding and often saturation.

\subsection{Evolutionary status}

\begin{table*}
\begin{minipage}{\textwidth}
\begin{center}
\caption{Differential abundances of ROA\,5701 relative to, young B-type stars from Hunter (in preparation), where available, and supplemented by \citet{{kil92},{kil94}}$^\ast$, and the interstellar medium \citep{wil00}.}
\label{tab_ay}
\begin{tabular}{@{}lccccccccccc}
\hline
Element	& \multicolumn{2}{c}{Young B-type stars}	& \multicolumn{2}{c}{ISM}		\\	
	& This paper		& \citet{moe98}		& This paper	& \citet{moe98}		\\
\hline
C	& $<$ --1.4		& $<$ --2.1	& $<$ --1.9     & $<$ --2.5   \\ 
N	& --0.5			& --0.7		& --0.8	     	& --1.0    	\\
O	& --0.8			& --0.6		& --0.9	     	& --0.7 	\\ 
Mg	& --1.1			& $\cdots$	& --1.1	     	& $\cdots$	\\
Si	& --1.5			& --1.3		& --1.3	     	& --1.1 	  \\
Al	& $<$ --1.0$^\ast$	& $\cdots$	& --1.1	     	& $\cdots$	\\
S	& --1.5$^\ast$		& $\cdots$	& --1.4	     	& $\cdots$	\\
Fe 	& $\cdots$		& --2.7	 	& $\cdots$   	& --2.6  	\\
\hline

\end{tabular}
\end{center}
\end{minipage}
\end{table*}

Studies of post-AGB stars have shown that some of these objects exhibit abundance trends similar to that of interstellar gas, i.e., with CNO and S near solar and Mg, Si, Ca, etc following the low abundances of Fe \citep{van95}. Examples of such objects include HD\,46703 \citep{luc84} and HD\,56126 \citep{par92}. Grain formation in the circumstellar shells of stars during the AGB evolution has been suggested to explain the origin of such a `cleaned-up' photosphere. The premise is that during the AGB phase, in the cool atmosphere of the star, a circumstellar shell is formed in which dust condenses. Dust particles are removed by radiation pressure, while the `cleaned' gas is re-accreted onto the stellar surface \citep{{mat92},{wat92}}. 
Iron, and similar metals, have a high condensation temperature, so are more likely to condense onto dust grains. Elements such as C, N and O, with lower condensation temperatures, show no significant depletions and stay in the gas phase \citep{par92}. Hence, the remaining re-accreted gas forms an iron-poor atmosphere. 

\citet{moe98} suggest that ROA\,5701 has undergone this gas--dust separation as they found a very low Fe abundance, with N and O relatively undepleted. However, in this scenario one would expect Fe to have a similar depletion to Mg and Si, as these would also be subject to grain formation \citep{ued05}. From our work we find that Mg and Si are underabundant by [Mg/H] = --1.1 $\pm$ 0.2 dex and [Si/H]= --1.3 $\pm$ 0.3 dex (relative to the ISM; Table \ref{tab_ay}). However \citet{moe98} derive a Fe abundance of [Fe/H] = --2.6 dex (relative to the ISM) from {\it Hubble Space Telescope} ({\it HST}) spectra, which is more than 1.0 dex lower than the Mg and Si values we find. Also, we have derived an abundance for S of [S/H] = --1.4 $\pm$ 0.2 dex (relative to the ISM), so the Mg, Si and S values agree with each other and are consistent with the average metallicity of the cluster, namely [Fe/H] = --1.2 dex \citep{pio05}. This result would remain unchanged if we had adopted abundance estimates relative to those from young B-type stars. If grain formation has occured, one would expect S to be relatively undepleted and therefore not have an adundance pattern in agreement with those of Mg and Si, thus implying that, for ROA\,5701, there is no evidence of a `cleaned-up' photosphere. There are two possibilities to explain the abundance patterns seen. Either the iron abundance suggested by \citet{moe98} is incorrect, or it is indeed very low and an alternative mechanism is needed to explain the abundance results found.

One explanation for the low Fe abundance is s-processing, with Fe reduced via successive neutron capture of $^{56}$Fe. This scenario has been applied to H-deficient very hot post-AGB stars, e.g., PG1159 type star K1-16 \citep*{{her03},{wer05}}. However s-process enriched objects are known to cover a range of metallicities between [Fe/H] = --0.2 to --1.0 dex \citep{van03}, much greater than the value \citet{moe98} find for ROA\,5701, i.e., [Fe/H] = --2.7 dex. Also, this mechanism is normally observed in post-AGB stars which have C-rich circumstellar material, such as HD\,158616 \citep*{are01}. By contrast, we have found C to be deficient by [C/H] $<$ --1.4 dex and so s-processing seems unlikely. 
The low C abundance could be explained by hot bottom burning (HBB; \citealt{moo01}), as suggested for V453\,Oph \citep{der05}. In this, temperatures become large enough for CNO processing to begin at the base of the convective envelope, with $^{12}$C converted to $^{14}$N and moved to the surface. \citet{her05} suggests that HBB may have an important role at extremely low metallicities. Only Population {\sc i} stars of initial mass M $\geq$ 5 M$_{\sun}$, and Population {\sc ii} stars of M $\geq$ 3.5 M$_{\sun}$, experience HBB \citep{dan96}, and as the initial mass of ROA\,5701 is unknown, no conclusions can be made regarding this process. 

The Fe abundance pattern observed in ROA\,5701 does not appear to be plausible in terms of s-processing. For the abundance patterns of the other elements, \citet{moe98} note that it is not necessary to invoke dredge-up during the AGB phase to explain these, but a moderate third dredge-up cannot be ruled out. The processes of dredge-up are well known (see, for example, \citealt{ibe83}). Briefly, dredge-up occurs during the stars evolution as the products of hydrogen and helium burning are transported from the core to the surface, changing the surface abundances. 
At the first dredge-up, the surface $^{14}$N abundance is increased at the expense of $^{12}$C, while the $^{16}$O abundance sees little change. During the second dredge-up, hydrogen products are again moved to the surface, with $^{12}$C and $^{16}$O almost completely converted to $^{14}$N. 
The third dredge-up sees the products of the helium burning cycle, following the triple--$\alpha$ process, brought to the surface. As a result,  $^{12}$C is enhanced relative to $^{14}$N and $^{16}$O, $^{14}$N being formed during the hydrogen burning and $^{16}$O formed through $\alpha$--capture by carbon atoms. Also s-process elements are dredged-up to the surface, and so would provide evidence for the third dredge-up \citep{luc93}. Objects which have undergone the third dredge-up would, therefore, have C-rich material, for example K\,648 in the globular cluster M\,15 \citep*{rau02}. 

Hotter post-AGB objects display a deficiency in C, unlike cool post-AGB stars, which suggests that they must evolve off the AGB before the third dredge-up begins, as the final convection would bring the products of helium burning to the surface \citep{moo02}. 
Carbon deficiencies have been observed in hotter objects with a lack of third dredge-up suggested, for example, ZNG--1 in the globular cluster M\,10 \citep{moo04}, Barnard 29 in M\,13 \citep{con94} and HD\,341617 \citep{moo02}. 

ROA\,5701 follows the trends seen in typical hot, post-AGB stars. The C abundance is [C/H] $<$ --1.4 $\pm$ 0.2 dex, with O and N underabundant by [O/H] = --0.8 $\pm$ 0.1 dex and [N/H] = --0.5 $\pm$ 0.2 dex, relative to young B-type stars (Table \ref{tab_ay}). Hence N, and O to a lesser extent, are enhanced relative to the other elements and the average metallicity of the cluster. \citet{moe98} observe a similar enhancement of N and O relative to the other elements. However their N abundance is not increased relative to O probably because they observed fewer N\,{\sc ii} lines. For C, only an upper limit could be set for the abundance, which indeed is lower than the average metallicity of the cluster. The C, N and O abundances measured here imply that ROA\,5701 has undergone the first and second dredge-ups as the hydrogen burning products appear to have been brought to the surface \citep{{mcc92},{con94}}. 
Therefore, the C underabundance suggests that, as with other post-AGB stars, ROA\,5701 has left the AGB before the onset of the third dredge-up. 

We have investigated whether ROA\,5701 could be an AGB--manqu\'{e} star originating from the Extreme Horizontal Branch (EHB), with diffusion still active and causing the low iron abundance observed by \citet{moe98}. \citet*{dor93} uses the terminology EHB to describe HB sequences of constant mass that do not reach the thermally pulsing stage on the AGB. Instead, following core helium exhaustion, they evolve into either post early--AGB stars, which leave the AGB before thermal pulsing, or AGB--manqu\'{e} stars, which never develop extensive outer convective zones and so do not reach the AGB. EHB stars have envelope masses that are too small to reach the end stages of normal AGB evolution. 
\citet{lan92} give the luminosity of ROA\,5701 as log (L/L$_{\sun}$) = 3.2, which combined with the evidence for the occurance of the first and second dredge-ups implies that ROA\,5701 has evolved along the AGB. Therefore, ROA\,5701 does not appear to be an AGB--manqu\'{e} star, so the low iron abundance cannot be explained via this scenario. 
However, it is reasonable that ROA\,5701 could be a post early--AGB star as there is no evidence for the occurance of the third dredge-up, see for example PHL\,1580 and PHL\,174 \citep{con91}, and ID6 in NGC\,5986 \citep{jas04}.

Assuming that the star has not undergone the third dredge-up or followed a gas--dust separation, the origin of the low Fe abundance suggested by \citet{moe98} remains unclear. Hence a more detailed investigation into this abundance is required. As stated previously, it is possible that with higher signal-to-noise optical spectra Fe\,{\sc iii} lines could be detected. We plan to seek such observational data in the future and examine the {\it HST}-UV spectra to improve upon the exisiting iron abundance estimate.

\section*{acknowledgments}

We are grateful to David Pinfield, Fergal Crawford, and the staff of the AAT for obtaining the observations. We would like to thank Ian Hunter for his assistance with data reduction and analysis, and Carrie Trundle for her help with line identification. 
We thank the anonymous referee for their helpful comments.
H.M.A.T. acknowledge financial support from the Northern Ireland Department of Education and Learning (DEL). 
F.P.K. is grateful to AWE Aldermaston for the award of a William Penney Fellowship. 

{}

\label{lastpage}


\begin{thebibliography}{}

\bibitem[\protect\citeauthoryear{Arellano Ferro, Giridhar \& Mathias}{Arellano et al.}{2001}]{are01} Arellano Ferro A., Giridhar S., Mathias P., 2001, A\&A, 368, 250

\bibitem[\protect\citeauthoryear{Barnes}{1993}]{bar93} 
	Barnes J., 1993, A Beginner's Guide to Using {\sc iraf}, NOAO Laboratory

\bibitem[\protect\citeauthoryear{Bedin et al.}{2004}]{bed04} Bedin L. R., Piotto G., Anderson J., Cassisi S., King I. R., Momany Y., Carraro G., 2004, ApJ, 605, L125

\bibitem[\protect\citeauthoryear{Cacciari et al.}{1984}]{cac84} Cacciari C., Caloi V., Castellani V., Fusi Pacci F., 1984, A\&A, 139, 285

\bibitem[\protect\citeauthoryear{Conlon et al.}{1991}]{con91} Conlon E. S., Dufton P. L., Keenan F. P., McCausland R. J. H., 1991, MNRAS, 248, 820

\bibitem[\protect\citeauthoryear{Conlon, Dufton \& Keenan}{Conlon et al.}{1994}]{con94} Conlon E. S., Dufton P. L., Keenan F. P., 1994, A\&A, 290, 897

\bibitem[\protect\citeauthoryear{D'Antona \& Mazzitelli}{1996}]{dan96} D'Antona F., Mazzitelli I., 1996, ApJ, 470, 1093

\bibitem[\protect\citeauthoryear{Deroo et al.}{2005}]{der05} Deroo P., Reyniers M., Van Winckel H., Goriely S., Siess L., 2005, A\&A, 438, 987

\bibitem[\protect\citeauthoryear{Dixon, Brown \& Landsman}{Dixon et al.}{2004}]{dix04} Dixon W. V., Brown T. M., Landsman W. B., 2004, ApJ, 600, L43

\bibitem[\protect\citeauthoryear{Dorman, Rood \& O'Connell}{Dorman et al.}{1993}]{dor93} Dorman B., Rood R. T., O'Connell R. W., 1993, ApJ, 419, 596

\bibitem[\protect\citeauthoryear{Dufton et al.}{1990}]{duf90} Dufton P. L., Brown P. J. F., Fitzsimmons A., Lennon D. J., 1990, A\&A, 232, 431

\bibitem[\protect\citeauthoryear{Dufton et al.}{2005}]{duf05} Dufton P. L., Ryans R. S. I., Trundle C., Lennon D. J., Hubeny I., Lanz T., Allende Prieto C., 2005, A\&A, 434, 1125

\bibitem[\protect\citeauthoryear{Eber \& Butler}{1988}]{ebe88} Eber F., Butler K., 1988, A\&A, 202, 153

\bibitem[\protect\citeauthoryear{Gies \& Lambert}{1992}]{gie92} Gies D. R., Lambert D. L., 1992, ApJ, 387, 673

\bibitem[\protect\citeauthoryear{Hambly et al.}{1996a}]{ham96a} Hambly N. C., Dufton P. L., Keenan F. P., Lumsden S. L., 1996a, MNRAS, 278, 811

\bibitem[\protect\citeauthoryear{Hambly et al.}{1996b}]{ham96b} Hambly N. C., Keenan F. P., Dufton P. L., Brown P. J. F., Saffer R. A., Peterson R. C., 1996b, ApJ, 466, 1018

\bibitem[\protect\citeauthoryear{Herwig}{2005}]{her05} Herwig F., 2005, ARA\&A, 43, 435

\bibitem[\protect\citeauthoryear{Herwig, Lugaro \& Werner}{Herwig et al.}{2003}]{her03} Herwig F., Lugaro M., Werner K., 2003, in Canberra, Kwok S., Dopita M., Sutherland R., eds, Proc. IAU Symp. 209, Planetary Nebulae: Their Evolution and Role in the Universe. Astron. Soc. Pac., p. 85

\bibitem[\protect\citeauthoryear{Hilker et al.}{2004}]{hil04} Hilker M., Kayser A., Richtler T., Willemsen P., 2004, A\&A, 422, L9

\bibitem[\protect\citeauthoryear{Howarth et al.}{2004}]{how04} Howarth I. D., Murray J., Mills D., Berry D. S., 2004, Starlink User Note 50.24: {\sc dipso} -- A Friendly Spectrum Analysis Program, Rutherford Appleton Laboratory/CCLRC

\bibitem[\protect\citeauthoryear{Hubeny}{1988}]{hub88} Hubeny I., 1988, Comp. Phys. Comm., 52, 103

\bibitem[\protect\citeauthoryear{Hubeny \& Lanz}{1995}]{hub95} Hubeny I., Lanz T., 1995, ApJ, 439, 875

\bibitem[\protect\citeauthoryear{Hubeny, Heap \& Lanz}{Hubeny et al.}{1998}]{hub98} Hubeny, I., Heap, S.R., Lanz, T., 1998, in Boulder--Munich, Howarth I. D., eds, ASP Conf. Ser. Vol. 131., Properties of Hot, Luminous Stars. Astron. Soc. Pac., San Francisco, p. 108

\bibitem[\protect\citeauthoryear{Hunter et al.}{2005}]{hun05} Hunter I., Dufton P. L., Ryans R. S. I., Lennon D. J., Rolleston W. R. J., Hubeny I., Lanz T., 2005, A\&A, 436, 687

\bibitem[\protect\citeauthoryear{Iben \& Renzini}{1983}]{ibe83} Iben I. Jr., Renzini A., 1983, ARA\&A, 21, 271

\bibitem[\protect\citeauthoryear{Jasniewicz et al.}{2004}]{jas04} Jasniewicz G., de Laverny P., Parthasarathy M., L\`{e}bre A., Th\'{e}venin F., 2004, A\&A, 423, 353

\bibitem[\protect\citeauthoryear{Kilian}{1992}]{kil92} Kilian J., 1992, A\&A, 262, 171

\bibitem[\protect\citeauthoryear{Kilian}{1994}]{kil94} Kilian J., 1994, A\&A, 282, 867

\bibitem[\protect\citeauthoryear{Kwok}{1993}]{kwo93} Kwok S., ARA\&A, 1993, 31, 63

\bibitem[\protect\citeauthoryear{Landsman et al.}{1992}]{lan92} Landsman W. B. et al., 1992, ApJ, 395, L21

\bibitem[\protect\citeauthoryear{Landsman et al.}{2000}]{lan00} Landsman W., Moehler S., Napiwotzki R., Heber U., Sweigart A., Catelan M., Stecher T., 2000, in Leige, Noels A., Magain P., Caro D., Jehin E., Parmentier G., Thoul A. A., eds, 35th Leige Int. Astrophys. Colloq., The Galactic Halo: From Globular Cluster to Field Stars. Inst. d'Astrop. et de Geop., Belgium, p. 515 

\bibitem[\protect\citeauthoryear{Lee et al.}{2005}]{lee05} Lee J -K., Rolleston W. R. J., Dufton P. L., Ryans R. S. I., 2005, A\&A, 429, 1025

\bibitem[\protect\citeauthoryear{Luck}{1993}]{luc93} Luck R. E., 1993, in Cambridge--Massachusetts, Sasselov D. D., eds, ASP Conf. Ser. Vol. 45, Luminous High--Latitude Stars. Astron. Soc. Pac., San Francisco, p. 87

\bibitem[\protect\citeauthoryear{Luck \& Bond}{1984}]{luc84} Luck R. E., Bond H. E., 1984, ApJ, 279, 729

\bibitem[\protect\citeauthoryear{Mathis \& Lamers}{1992}]{mat92} Mathis J. S., Lamers H. J. G. L. M., 1992, A\&A, 259, L39

\bibitem[\protect\citeauthoryear{McCausland et al.}{1992}]{mcc92} McCausland R. J. H., Conlon E. S., Dufton P. L., Keenan F. P., 1992, ApJ, 394, 298 

\bibitem[\protect\citeauthoryear{Massey}{1997}]{mas97} Massey P., 1997, A User's Guide to CCD Reductions with {\sc iraf}, NOAO Laboratory

\bibitem[\protect\citeauthoryear{Massey, Valdes \& Barnes}{Massey et al}{1992}]{mas92} Massey P., Valdes F., Barnes J., 1992, A User's Guide to Reducing Slit Spectra with {\sc iraf}, NOAO Laboratory

\bibitem[\protect\citeauthoryear{Moehler}{2001}]{moe01} Moehler S., 2001, PASP, 113, 1162

\bibitem[\protect\citeauthoryear{Moehler et al.}{1998}]{moe98} Moehler S., Heber U., Lemke M., Napiwotzki R., 1998, A\&A, 339, 537

\bibitem[\protect\citeauthoryear{Mooney et al.}{2001}]{moo01} Mooney C. J., Rolleston W. R. J., Keenan F. P., Dufton P. L., Pollacco D. L., Magee H. R., 2001, MNRAS, 326, 1101

\bibitem[\protect\citeauthoryear{Mooney et al.}{2002}]{moo02} Mooney C. J., Rolleston W. R. J., Keenan F. P., Dufton P. L., Smoker J. V., Ryans R. S. I., Aller L. H., 2002, MNRAS, 337, 851

\bibitem[\protect\citeauthoryear{Mooney et al.}{2004}]{moo04} Mooney C. J., Rolleston W. R. J., Keenan F. P., Dufton P. L., Smoker J. V., Ryans R. S. I., Aller L. H., Trundle C., 2004, A\&A, 419, 1123

\bibitem[\protect\citeauthoryear{Napiwotzki, Heber \& K\"{o}ppen}{Napiwotzki et al.}{1994}]{nap94} Napiwotzki R., Heber U., K\"{o}ppen J., 1994, A\&A, 292, 239
	
\bibitem[\protect\citeauthoryear{Norris}{1974}]{nor74} Norris J., 1974, ApJ, 194, 109

\bibitem[\protect\citeauthoryear{Norris}{2004}]{nor04} Norris J., 2004, ApJ, 612, L25

\bibitem[\protect\citeauthoryear{Oudmaijer}{1996}]{oud96} Oudmaijer R. D., 1996, A\&A, 306, 823

\bibitem[\protect\citeauthoryear{Parthasarathy, Garcia Lario \& Pottasch}{Parthasarathy et al.}{1992}]{par92} Parthasarathy M., Garcia Lario P., Pottasch S. R., 1992, A\&A, 264, 159

\bibitem[\protect\citeauthoryear{Piotto et al.}{2005}]{pio05} Piotto et al., 2005, ApJ, 621, 777 

\bibitem[\protect\citeauthoryear{Rauch, Heber \& Werner}{Rauch et al.}{2002}]{rau02} Rauch T., Heber U., Werner K., 2002, A\&A, 381, 1007

\bibitem[\protect\citeauthoryear{Ryans et al.}{1996}]{rya96} Ryans R. S. I., Hambly N. C., Dufton P. L., Keenan F. P., 1996, MNRAS, 278, 132

\bibitem[\protect\citeauthoryear{Ryans et al.}{2003}]{rya03} Ryans R. S. I., Dufton P. L., Mooney C. J., Rolleston W. R. J., Keenan F. P., Hubeny I., Lanz T., 2003, A\&A, 401, 1119

\bibitem[\protect\citeauthoryear{Sarkar, Parthasarathy \& Reddy}{Sarkar et al.}{2005}]{sar05} Sarkar G., Parthasarathy M., Reddy B. E., 2005, A\&A, 431, 1007

\bibitem[\protect\citeauthoryear{Sigut}{1996}]{sig96} Sigut T. A. A., 1996, ApJ, 473, 452

\bibitem[\protect\citeauthoryear{Trundle et al.}{2004}]{tru04} Trundle C., Lennon D. J., Puls J., Dufton P. L., 2004, A\&A, 417, 217

\bibitem[\protect\citeauthoryear{Ueda et al.}{2005}]{ued05} Ueda Y., Mitsuda K., Murakami H., Matsushita K., 2005, ApJ, 620, 274

\bibitem[\protect\citeauthoryear{Valdes}{1993}]{val93} Valdes F., 1993, Guide to the Slit Spectra Reduction Task {\sc doecslit}, NOAO Laboratory

\bibitem[\protect\citeauthoryear{Van Winckel}{2003}]{van03} Van Winckel H., 2003, ARA\&A, 41, 391

\bibitem[\protect\citeauthoryear{Van Winckel, Waelkens \& Waters}{Van Winckel et al.}{1995}]{van95} Van Winckel H., Waelkens C., Waters L. B. F. M., 1995, A\&A, 293, L25

\bibitem[\protect\citeauthoryear{Wallace \& Clayton}{1996}]{wal96} Wallace P. T., Clayton C. A., 1996, Starlink User Note 78.8: {\sc rv} -- Radial Components of Observer's Velocity, Rutherford Appleton Laboratory/CCLRC

\bibitem[\protect\citeauthoryear{Waters, Trams \& Waelkens}{Waters et al.}{1992}]{wat92} Waters L. B. F. M., Trams N. R., Waelkens C., 1992, A\&A, 262, L37

\bibitem[\protect\citeauthoryear{Werner \& Herwig}{2005}]{wer05} Werner K., Herwig F., 2006, PASP, 118, 183

\bibitem[\protect\citeauthoryear{Wilms, Allen \& McCray}{Wilms et al.}{2000}]{wil00} Wilms J., Allen A., McCray R., 2000, ApJ, 542, 914

\bibitem[\protect\citeauthoryear{Zijlstra et al.}{1992}]{zij92} Zijlstra A. A., Loup C., Waters L. B. F. M., de Jong T., 1992, A\&A, 265, L5

\bibitem[\protect\citeauthoryear{Zinn, Newell \& Gibson}{Zinn et al.}{1972}]{zin72} Zinn R. J., Newell E. B., Gibson J. B., 1972, A\&A, 18, 390



\end{thebibliography}
\end{document}